# Adaptive Wavelets Applied to the Analysis of Nonlinear Systems with Chaotic Dynamics

**V. A. Gusev, A. E. Hramov*, and A. A. Koronovskiĭ**
*Saratov State University, Saratov, Russia*

**e-mail: aeh@cas.ssu.runnet.ru*

**Abstract**—We consider an approach to the analysis of nonstationary processes based on the application of wavelet basis sets constructed using segments of the analyzed time series. The proposed method is applied to the analysis of time series generated by a nonlinear system with and without noise.

As is known, the analysis of nonlinear dynamic systems of various natures featuring complicated oscillatory regimes requires special methods [1–4]. Among these methods, of special interest is the wavelet analysis [5–7] offering a powerful means of diagnostics of the behavior of nonlinear dynamic systems [5, 8, 9]. Presently, wavelets are successfully applied to the analysis of nonstationary signals generated by systems of biological and medical nature [10, 11], geophysical processes [5, 12], electron-plasma systems [4, 13, 14], chaotic and turbulent processes [7, 15–17], etc. An important modification of the wavelet analysis consists in using specially constructed (adaptive) wavelet basis sets, which allows the analysis to reveal certain features of the analyzed signals [8, 18].

In this paper, we describe a new approach to the analysis of time series generated by nonlinear processes. The proposed method is based on the use of adaptive wavelet basis sets constructed using segments of the analyzed time series, which facilitates the separation of characteristic features and structures from signals (even in the presence of noise).

The continuous wavelet transform of a function $x(t)$ is defined as

$$W(t, s) = \frac{1}{\sqrt{s}} \int_{-\infty}^{+\infty} x(t')\psi^*\left(\frac{t-t'}{s}\right)dt', \qquad (1)$$

where $s$ is the analyzed time scale and $\psi$ is the base wavelet function (the asterisk denotes complex conjugation). The latter function, albeit it can be selected rather arbitrarily, must satisfy certain requirements, the most important of which are as follows. First, the base wavelet function should be localized in both time and space representations; second, this function has to obey the condition of zero mean $\int_{-\infty}^{+\infty}\psi(t)dt = 0$ or an equivalent relation $\hat{\psi}(0) = 0$, where $\hat{\psi}$ is the Fourier transform of the base wavelet function.

Let us assume that the signal $x(t)$ is generated by a nonlinear dynamic system occurring in an oscillatory regime (in particular, in the state of dynamic chaos). Consider the following procedure for constructing a complex wavelet function $\psi$ using the given time series $x(t)$.

First, we separate a characteristic time scale $\tau$ (with the corresponding frequency $\omega_\tau = 2\pi/\tau$ of the nonstationary process $x(t)$. In the case of periodic oscillations, $\tau$ coincides with the period. For a chaotic nonregular signal, $\tau$ can be defined by various means. For example, if the chaotic signal $x(t)$ is characterized by the phase $\Phi$ determined using the Hilbert transform [2], the characteristic frequency can be defined as $\omega_\tau = \lim_{t\to\infty}\Phi(t)/t$. In the simplest case, the frequency $\omega_\tau$ can be determined as corresponding to the most intense peak in the Fourier spectrum of power $P(\omega)$ of the signal $x(t)$.

Second, we select a certain initial moment $t_0$ in the time series $x(t)$, relative to which the wavelet basis set will be constructed. The real and imaginary parts of the base wavelet function $\psi$ are constructed according to the formulas

$$\mathrm{Re}\psi(t) = \pi^{-1/4}\left\langle x(t-t_0)\exp\left(-\frac{1}{2}\frac{(t-t_0)^p}{(n\tau)^p}\right)\right\rangle,$$
$$\mathrm{Im}\psi(t) = \pi^{-1/4}\left\langle x(t-t_0+\tau/2)\exp\left(-\frac{1}{2}\frac{(t-t_0)^p}{(n\tau)^p}\right)\right\rangle, \qquad (2)$$

where $\langle\cdot\rangle$ denotes the operation of elimination of the mean value for satisfying the zero mean condition; $n$ and $p$ are the wavelet parameters. As can be seen from formula (2), the wavelet parameter $t_0$ characterizes the segment of series $x(t)$ which most significantly influences the wavelet spectrum $W(t, s)$. Note that the base wavelet function of type (2) for $x(t) = \sin\omega_\tau t$ with $n = 1.0$ and $p = 2.0$ is an analog of the widely used Morlet wavelet [5, 8] that meets all requirements to the wavelet functions. For a base wavelet function determined in the form (2), the quantity $f_s = 1/s$ has a meaning analogous to the frequency of the Fourier transform [7, 8].

Using a wavelet transform with the base function of type (2), it is possible to effectively separate segments with the length $\Delta t \approx 4n\tau$ from the time series $x(t)$, which are "like" the segment where $t \in (t_0 - 2n\tau, t_0 + 2n\tau)$. Note that the amplitude of $\psi(t)$ decreases by a factor of $\exp(2^{(p-1)})$ within the time interval $|t - t_0| = 2n\tau$, which implies that the wavelet function is based on a segment of the time series with the middle at $t = t_0$ and a length of $4n\tau$.

For illustrating the above procedure, let us consider the results of the wavelet analysis of a signal generated by the Rössler system [1]

$$\dot{x} = -(y + z), \quad \dot{y} = x + ey, \quad \dot{z} = \omega - mz + xz, \quad (3)$$

which represents a standard nonlinear flow systems featuring chaotic dynamics. Below, we consider the system of type (3) with the control parameters $e = w = 0.2$ and $m = 5.8$, which corresponds to the case of a ribbon chaos.

Figure 1a shows the time series $x(t)$, while Fig. 1b presents the results of calculation of the wavelet spectrum $|W(t, f_s)|$ for a wavelet function constructed using the segment of $x(t)$ corresponding to $t_0 = 164.5$ ($n = 1.0$, $p = 8.0$) (in Fig. 1a, this segment is indicated by vertical dashed lines). The spectrum in Fig. 1b is plotted as the frequencies $f_s$ (corresponding to the scales $s$ of the wavelet transform) versus time $t$. The shape of this spectrum indicates that the time scale $s = \tau$ is always present, but the intensity varies with the time $t$.

The maxima of the wavelet spectrum $|W(t, s \approx \tau)|$ taking place at the moments $t = \hat{t}$ correspond to the regions of the time series which are close to the segment used for constructing the base wavelet function $\psi$ (2). This implies that the integral $\int_{t-2n\tau}^{t+2n\tau} [x(t + t') - (t_0 + t')]^2 dt$ exhibits minima at $t \sim \hat{t}$, as illustrated in Fig. 1c showing the segments of time series corresponding to the time $\hat{t} = 210$ (indicated by arrow "c" in Fig. 1b), at which the wavelet surface exhibits a maximum. For the comparison, Fig. 1d shows the segments of time series corresponding to the time $\hat{t} = 133$ (indicated by arrow "d" in Fig. 1b), at which the wavelet spectrum $|W(t, s \approx \tau)|$exhibits a maximum. Dashed curves in Figs. 1c and 1d show the initial segments of the time series used for the formation of the base wavelet function $\psi$. A comparison of Figs. 1c and 1d shows that the segments of the time series virtually coincide in the former case and exhibit significant distinctions in the latter case.

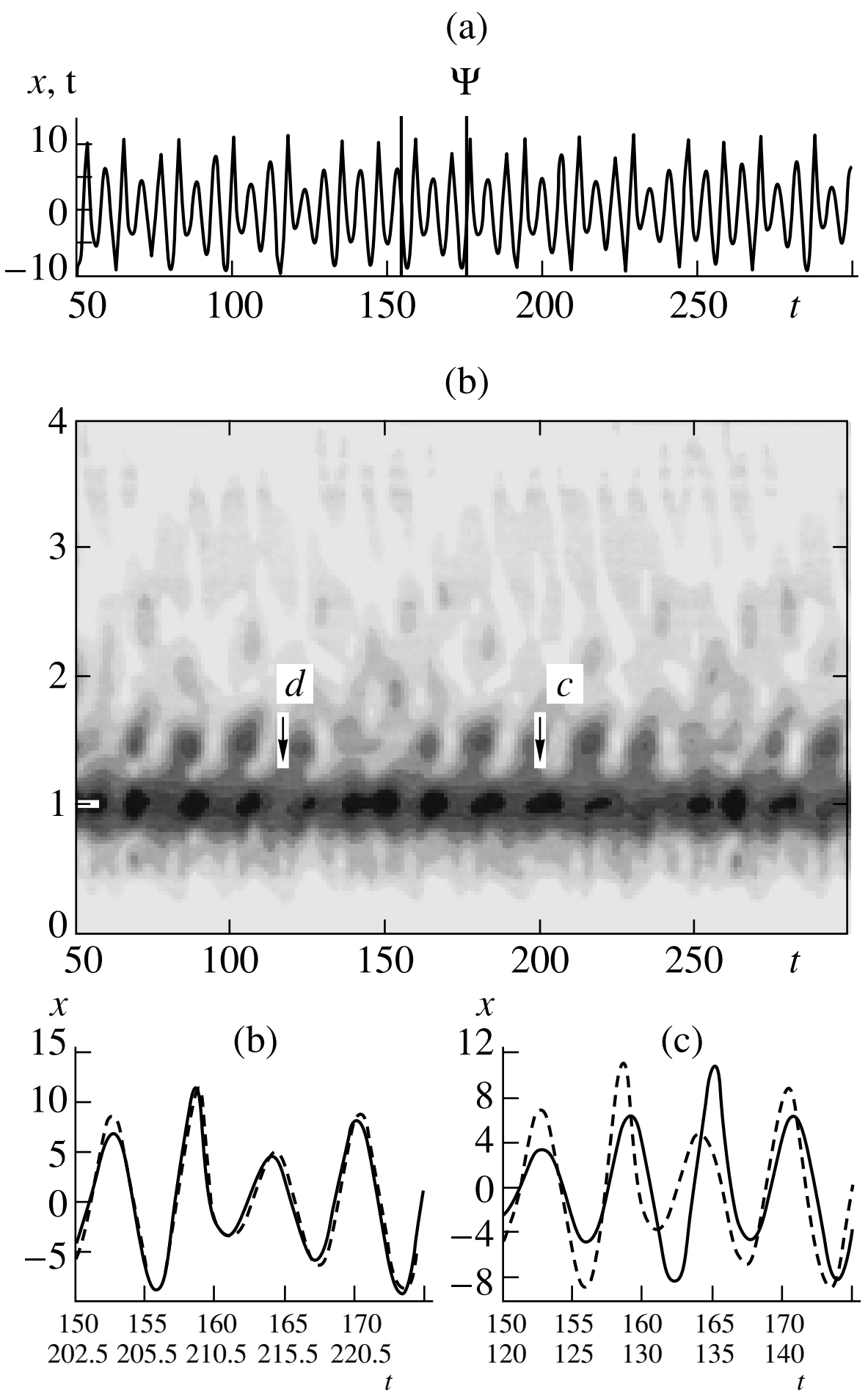

**Fig. 1.** Wavelet analysis of a noisless signal: (a) time series $x(t)$ generated by the Rössler system in a chaotic regime; (b) the corresponding wavelet spectrum $|W(t, f_s)|$; ($c$, $d$) comparison of the segments of the time series $x(t)$ corresponding to the maximum and minimum of the wavelet spectrum (indicated by arrows in (b)). Symbol $\psi$ in (a) and the dashed curves in (c, d) indicates a segment used for constructing the base wavelet function. In (c, d), the first row of values on the abscissa axis refers to the base function (dashed curves) and the second row, to the compared segment (solid curve) of the initial time series. 1

Real signals are usually distorted by noise and an important problem consists in finding a means of effectively separating the useful signal $x(t)$ from the initial noisy time series. Let us consider a time series $y(t)$ comprising a superposition of the deterministic signal $x(t)$ and a noise component:

$$y(t) = x(t) + D\xi(t), \quad (4)$$

where $\xi(t)$ is a random function modeling the Gaussian white noise. As above, we assume that the deterministic signal $x(t)$ is generated by the Rössler system (3) with the same control parameters. Consider the wavelet spectra of the signal $y(t)$ (4) obtained for various noise amplitudes $D$ (i.e., various signal to noise ratios $\chi$ =

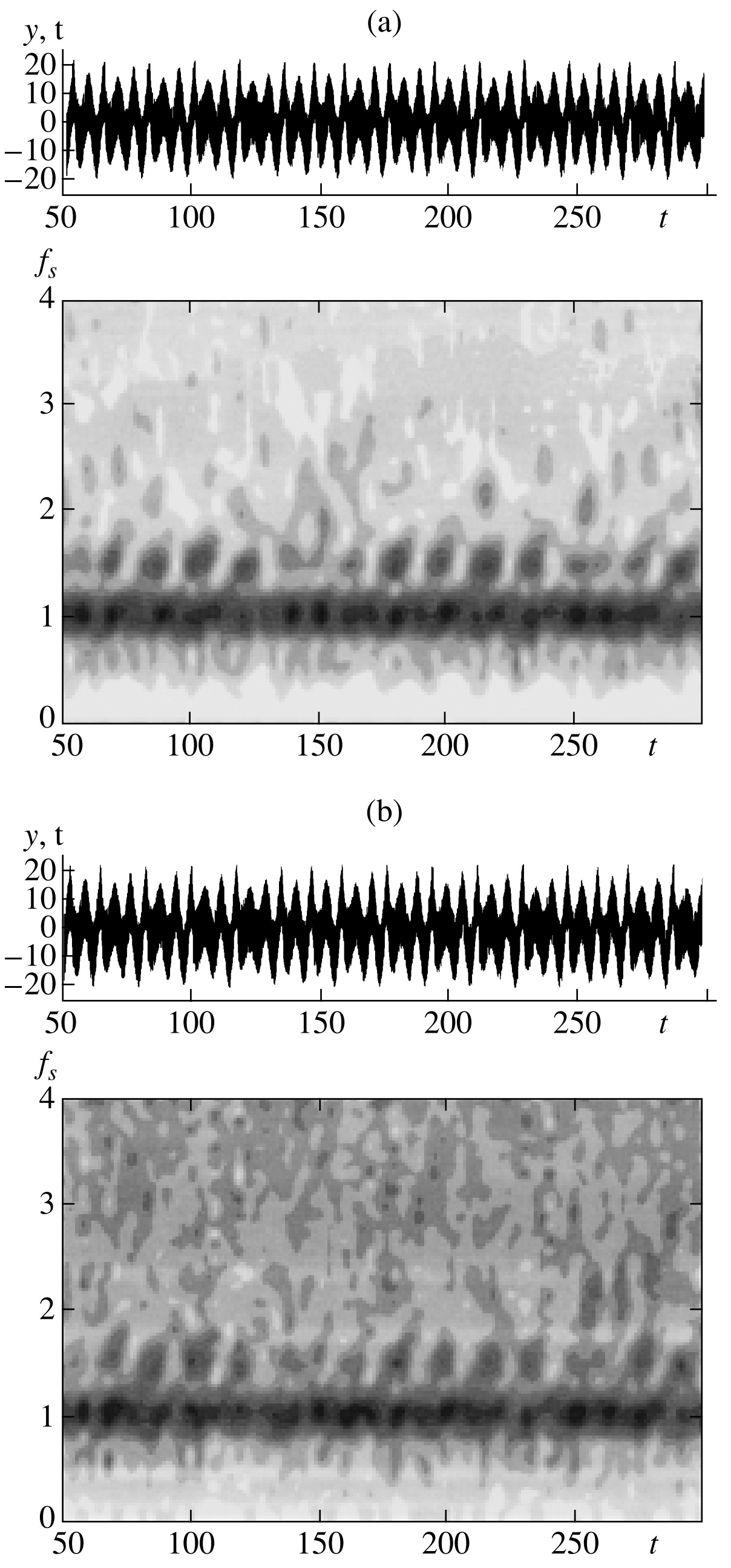

**Fig. 2.** Wavelet analysis of noisy signals: time series $y(t)$ and the corresponding wavelet spectra for different relative noise intensities $\chi$ = 1.0 (a) and 0.5 (b). The base wavelet function was constructed using the same segment of the time series as in Fig. 1.

$|x_{max}|/D$, where $|x_{max}|$ is the maximum value of variable $x(t)$).

An analysis of the wavelet spectra constructed for the time series (4) with different noise amplitudes showed that effective separation of the characteristic features of the deterministic signal $x(t)$ by means of the wavelet transform is possible for the signal to noise ratio $\chi > 0.2$–$0.3$. This is illustrated by Fig. 2 showing the time series $y(t)$ of type (4) and the corresponding wavelet spectra $|W(t, f_s)|$ constructed for two values of the signal to noise ratio, $\chi = 1.0$ (Fig. 2a) and 0.5 (Fig. 2b). For the convenience of comparison of the results of separating the useful signal $x(t)$ from the noise, the time series and wavelet spectra in Fig. 2 are constructed for the same time intervals as in Fig. 1. The base wavelet function (2) was constructed using a segment of the noisy time series with $t_0 = 164.5$ for $n = 1.0$ and $p = 8.0$ (cf. Fig. 1a).

As can be seen from Fig. 2 (with reference to Fig. 1 constructed in the absence of noise), the wavelet transform with a base function (2) representing a segment of the analyzed noisy time series (as well as in the case of the signal free of a noise component) allows us to separate and analyze the characteristic features of the useful signal. For a relatively small noise intensity (Fig. 2a, $\chi = 1.0$), the form of the resulting wavelet surface only slightly differs from the corresponding spectrum of the signal without noise (Fig. 1b). As the noise intensity increases (Fig. 2b, $\chi = 0.5$), the form of the wavelet surface is distorted and the maxima are less pronounced. The wavelet spectrum in Fig. 2b is especially strongly impaired (in comparison to Fig. 1b) in the region of scales $f_s > 2.0$. At the same time, the structure of the wavelet surface $|W(t, f_s)|$ in the region of $s \sim 1.0$ is close to that of the spectrum obtained for the signal without noise, which provides for a correct analysis. As the noise intensity grows further, the wavelet surface structure is broken even in the region of the time scales $s \sim 1.0$, which hinders separation and analysis of the initial signal $x(t)$ in the presence of noise.

Thus, we have demonstrated the possibility of effective analysis of the time series (including noisy signals) by means of the wavelet analysis using basis functions representing segments of the analyzed time series.

**Acknowledgments.** This study was supported by the Russian Foundation for Basic Research, project nos. 02-02-16351 and 01-02- 17392.